# Multi-Flow Attacks Against Network Flow Watermarks: Analysis and Countermeasures

Negar Kiyavash, *Member, IEEE,* Amir Houmansadr, *Student Member, IEEE,* and Nikita Borisov, *Member, IEEE*

**Abstract**—In this paper, we analyze several recent schemes for watermarking network flows that are based on splitting the flow into timing intervals. We show that this approach creates time-dependent correlations that enable an attack that combines multiple watermarked flows. Such an attack can easily be mounted in nearly all applications of network flow watermarking, both in anonymous communication and stepping stone detection. The attack can be used to detect the presence of a watermark, recover the secret parameters, and remove the watermark from a flow. The attack can be effective even if different flows are marked with different values of a watermark.
We analyze the efficacy of our attack using a probabilistic model and a Markov-Modulated Poisson Process (MMPP) model of interactive traffic. We also implement our attack and test it using both synthetic and real-world traces, showing that our attack is effective with as few as 10 watermarked flows. Finally, we propose possible countermeasures to defeat the multi-flow attack.

## 1 Introduction

Traffic analysis is the practice of inferring sensitive information from communication patterns. Traffic analysis has been particularly studied in the context of anonymous communication systems, where features such as packet timings, sizes, and counts can be used to link two flows and break anonymity guarantees [1], [2]. Traffic analysis is also sometimes used in intrusion detection, for example, to detect the presence of stepping stones within an enterprise [3].

Recently, there has been a growing interest in the use of *watermarking* to aid traffic analysis [4], [5], [6], [7], [8], [9], [10]. In this case traffic patterns of a flow (usually packet timings) are actively modified to contain a special pattern. If the same pattern is later found on another flow, the two are easily linked. Watermarking significantly reduces the computation and communication costs of traffic analysis, and may also lead to more precise detection with fewer false positives [9]. Watermarking has been applied to both the problems of attacking anonymity systems [5], [7], [8] and detecting stepping stones [4], [6].

In both contexts, many flows must be watermarked in order to learn new information. In our work, we consider whether an attacker[1] can learn enough information to defeat the watermark by observing multiple watermarked flows [11]. We apply this multi-flow threat model to the latest generation of *interval-based watermarks* [6], [7], [8]. These watermarks subdivide the flow to be marked into discrete time intervals and perform transformative operations on an entire interval of packets. This approach is more robust to packet losses, insertions, and repacketization, than previous approaches that focused on individual packets [4], [5], because the time intervals allow the watermarker and detector to retain synchronization. However, the same synchronization property can be exploited by attackers by "lining up" multiple watermarked flows and observing the transformations that were inserted.

We show through experiments that the interval-based watermark schemes are completely vulnerable to an attacker who can collect a small number of watermarked flows—about 10. This is sufficient to not only detect that a watermark is indeed present,

• *N. Kiyavash, A. Houmansadr, and N. Borisov are with the University of Illinois at Urbana-Champaign, Urbana, IL, 61820.*
*E-mail: {kiyavash, ahouman2, nikita}@illinois.edu*

1. We use "attacker" here to refer to someone attacking the watermarking scheme; in the case where watermarks themselves are used by attackers, these will be the "counter-attackers."

but also to recover the secret parameters of the watermark scheme and to be able to remove the watermark at a low cost. Furthermore, our attack works even if different watermarked flows contain different embedded "messages" with only about twice the number of watermarked flows necessary. We also analytically estimate the false-positive rates for our attack and find them to be very low.

We also consider some countermeasures to such attacks. We show that by using multiple "keys" (time interval assignments) to watermark different flows, it is possible to defeat our attack. This countermeasure comes at a cost of higher computation overhead at the detector and a higher rate of false positives. However, this increased cost is only linear, whereas the increased cost of the attacker is superexponential, thus providing an effective defense.

The rest of the paper is organized as follows. The next section presents the setting for our attack and reviews the three schemes considered in this paper. Section 3 describes the theoretical foundation for our attack, and Section 4 implements the attack. We discuss potential countermeasures to the attack in Section 5. Section 6 concludes.

## 2 Background

We first describe the setting of our attack in a bit more detail and then review the essential details of the watermarking schemes we analyze.

### 2.1 Network Flow Watermarking

The setting for network flow watermarking is similar to that of other digital media watermarks (and in fact uses similar techniques). The general model, as shown in Figure 1, involves a network flow passing through a watermarking point (typically a router of some sort) which transforms or distorts the flow in some way (typically by modifying packet timings by adding artificial delays in forwarding). In the general setting, the watermarker has a secret *key* and uses it to encode a *message* in the traffic characteristics.

After watermarking, the flow undergoes some natural or intentional distortion. Natural distortion can take the form of delays at intermediate routers (or rather, variability of delays, i.e., *jitter*), but may also include dropped or retransmitted packets, repacketization, and other changes. In addition, an attacker may intentionally distort traffic characteristics in order to prevent the watermark from being recovered.

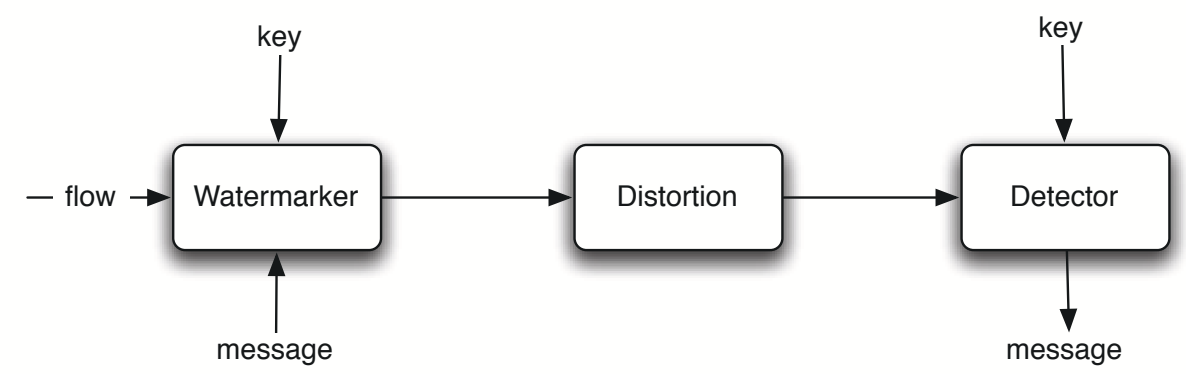

Fig. 1. Network Flow Watermarking

The distorted flow finally arrives at a detection point. The detector shares the secret key and uses it to extract the message encoded in the watermark. A good watermark will allow reliable recovery of the message from the watermarked flow despite the intermediate distortion.

In network flow watermarks, the *message* component of the watermark may be used in two ways. First, all watermarked flows may be marked with a single message. In this case, the detector's main goal is to decide whether the watermark is present or not by checking whether the decoded message is the correct one. Alternately, different flows may have a different message embedded, so that when a watermarked flow is detected, it can be linked with a particular marked flow. This comes at a cost of less reliable detection, since the single-message context creates more opportunities to detect errors. Our attacks are designed to work in both single-message and multiple-message contexts.

### 2.2 Watermarks in Anonymous Systems

At a very high level, an anonymous system maps a number of input flows to a number of output flows while hiding the relationship between them. The internal operation can be implemented by a mix network [12], onion routing [13], or a simple proxy [14]. The goal of an attacker, then, is to link an incoming flow to an outgoing flow (or vice versa).

A watermark can be used to defeat anonymity protection in low-latency anonymous systems by marking certain input flows and watching for marks on the output flows. For example, a malicious website might insert a watermark on all flows from the site to the anonymizing system. A cooperating attacker who can eavesdrop on the link between a user and the anonymous system can then determine if the user is browsing the site or not. Similarly, a compromised entry router in Tor [15] can watermark all of its flows, and cooperating exit routers or websites can detect this watermark.

Note that this does not enable a fundamentally new attack on low-latency anonymous systems: it has been long known [13] that if an attacker can observe a flow at two points, he can determine if the flow is the same, unless cover traffic is used. (In fact, deployed low-latency systems such as Onion Routing [13], Freedom [16], Tor [15], and AN.ON [17] have all opted to forego cover traffic due to it being expensive, hoping instead that it will be difficult for an attacker to observe a significant fraction of incoming and outgoing flows.) However, watermarking makes the attack much more efficient. With passive traffic analysis, if one attacker observes $n$ input flows and another observes $m$ output flows, the attack will require $O(n)$ communication between the attackers and $O(nm)$ computation, as one attacker must transmit characteristics of all $n$ flows to the other, and then each output flow must be matched against each input flow. With watermarking, on the other hand, no communication needs to take place between the two attackers after they have established a shared secret key, and the computation cost is $O(n)$ and $O(m)$ at the watermarker and detector respectively, as the watermarker marks each input flow and the detector checks each output flow for the presence of a mark.

Multi-Flow Attack (MFA): In the above examples, a website, or an input router, will insert the watermark into all the input flows going through them. Therefore, it will be possible for the anonymous system to obtain multiple watermarked flows. These flows can then be used to recover the secret key and then remove the watermarks from subsequent flows, using the techniques we describe below. Our techniques are low-cost, requiring a small number of watermarked flows and modest computation, so it is easy to check whether watermarking is being applied by a given website or router by aggregating its flows.

The only context where our attack does not apply is in a *traffic confirmation attack*. In this case, an attacker already has a strong suspicion that a particular input flow corresponds to a particular output flow, and therefore need only watermark a single flow. Traffic confirmation attacks are a more rare use of traffic analysis, since they only confirm existing suspicions, rather than revealing new linkages between flows. Furthermore, the efficiency gains of watermarks are not beneficial in this case, since $n = m = 1$. Therefore, our attack will apply to the vast majority of practical uses of watermarks in anonymous systems.

### 2.3 Watermarks in Stepping Stones

A stepping stone is a host that is used to relay traffic through an enterprise network to another remote destination, in order to hide the true origin of the flow. To detect such hosts, an enterprise must be able to link an incoming flow to the relayed outgoing flow. The situation is therefore very similar to an anonymous communication system, with $n$ flows entering the enterprise and $m$ flows leaving. Once again, this task may be accomplished by passive traffic analysis [3], [18], [19], [20], but watermarks make such detection much more efficient. Passive techniques will require $O(nm)$ computation and potentially $O(n)$ communication, if there are multiple border routers through which traffic can enter or leave the enterprise. With watermarking, border routers for an enterprise will insert watermarks on all incoming flows, and check for the presence of the mark on all outgoing flows, as shown in Figure 2, reducing the computation cost to $O(n)$ and $O(m)$ for the incoming and outgoing flows.

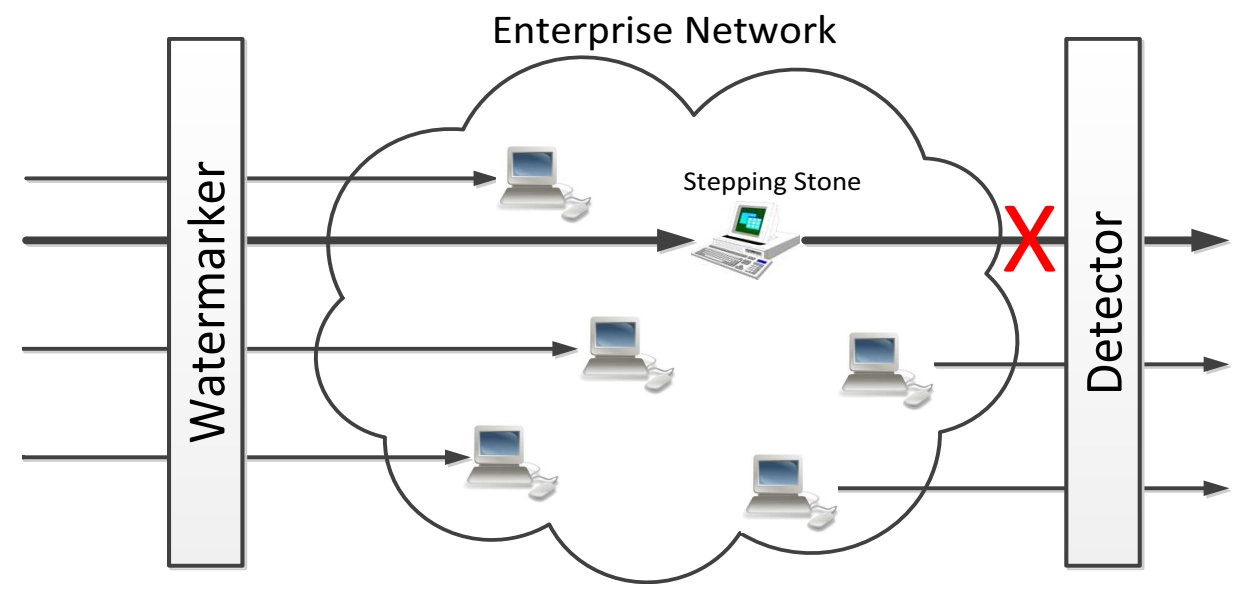

Fig. 2. Stepping stone detection architecture.

Multi-Flow Attack: Since all incoming flows must be marked, an attacker in control of a compromised host can simply generate multiple external flows destined for that host (and not relay them), and then collect the timing characteristics of the flows as they arrive at the host to recover the secret watermark key. Once this is accomplished, the key can be used to remove watermarks from relayed flows, thus defeating stepping stone detection.

### 2.4 Interval Centroid-based Watermarking (ICBW)

We next review the scheme proposed by Wang et al. [7]; for more details of the scheme as well as some analysis we refer the reader to [7]. The scheme is based on dividing the stream into intervals of equal lengths, using two parameters: $o$, the offset of the first interval, and $T$, the length of each interval. A subset of $2n$ of these intervals is randomly selected which is subsequently randomly divided into two further subsets $A$ and $B$ each consisting of $n = rl$ intervals. Each of the sets $A$ and $B$ are randomly divided to $l$ subsets denoted by $\{A_i\}_{i=1}^{l}$ and $\{B_i\}_{i=1}^{l}$ each consisting of $r$ intervals. The $i$-th watermark bit is encoded using the sets $\{A_i, B_i\}$. Therefore, a watermark of length $l$ can be embedded in the flow.

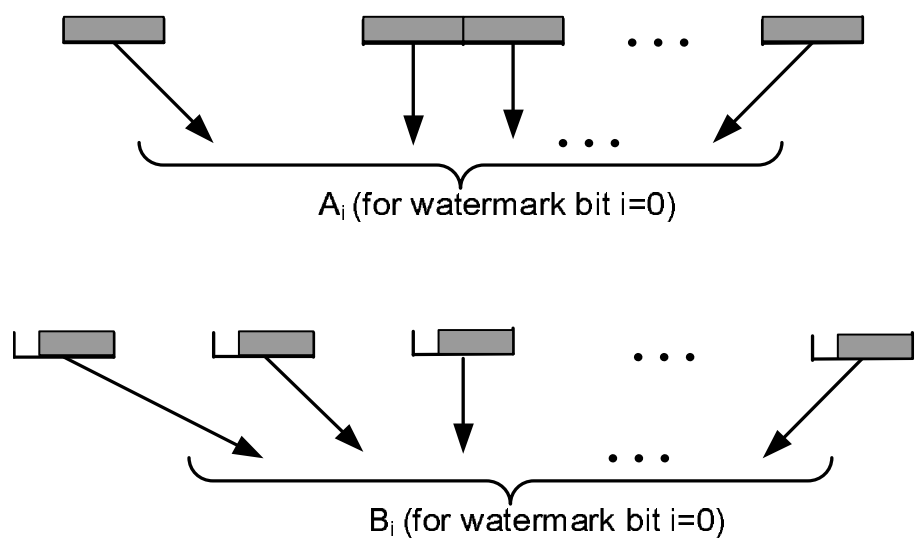

Fig. 3. Insertion of watermark bit 0 by ICBW.

The watermarker and detector agree on the parameters $o$, $T$ and use a pseudorandom number generator (PRNG) and a seed $s$ to randomly select and assign intervals for watermark insertion. To keep the watermark transparent, all of these parameters are kept secret. Depending on whether the $i$-th watermark bit is $1$ or $0$, the watermarker delays the arrival times of the packets at the interval positions in sets $A_i$ or $B_i$ respectively, by a maximum of $a$. Figure 3 shows insertion of a watermark bit of 0.

As the result of this embedding scheme, the expected value of aggregate centroid, i.e., the average of the arrival time of the packets modulo the length of the interval $T$, in either the intervals $A_i$ (when watermark bit is 1) or $B_i$ (when watermark bit is $0$) corresponding to bit $i$ is increased by $\frac{a}{2}$. The expected difference between the aggregate centroid of $A_i$ and $B_i$ now will be $\frac{a}{2}$ when watermark bit is $1$ or $-\frac{a}{2}$ when watermark bit is $0$.

The detector checks for the existence of the watermark bits. The check on watermark bit $i$ is performed by testing whether the difference of the aggregate centroid of packet arrival times in the intervals $A_i$ and $B_i$ is closer to $\frac{a}{2}$ or $-\frac{a}{2}$. If it is closer to $\frac{a}{2}$, then the watermark bit is decoded as $1$ and if it is closer to $-\frac{a}{2}$, the bit is declared a $0$. By focusing on the arrival times of many intervals ($r$ of them for each bit of watermark) rather than individual packet timings, ICBW approach is robust to repacketization, insertion of chaff, and mixing of data flows. Network jitter can shift packets from one interval into another, but the suggested parameters for $a$ and $T$ (350ms and 500ms respectively) are large enough that few packets will be affected.

The secrecy of the interval positions $A_i$ and $B_i$ make the mark difficult to detect or remove, as it is hard to distinguish the patterns generated by the mark from natural variation in traffic rates. We show in Sections 3 and 4, however, that a simple technique allows an observer to effectively recover the watermark positions and values. This technique is applicable to any watermarking scheme that creates periods of clear or low traffic at *specific* parts of the flows across many flows. Next, we briefly describe *Interval-Based Watermarking (IBW)*, a flow watermarking scheme proposed by Pyun et al. [6] to detect *stepping stones*. Our attacks also applies to this scheme.

### 2.5 Interval-Based Watermarking

Similar to ICBW, the watermarking scheme of Pyun et al. [6] manipulates the arrival times of the packets over a set of preselected intervals. The watermark embedding is achieved by manipulating the rates of traffic in successive intervals. There are two manipulations: an interval $I_i$ may be *cleared* by delaying all packets from interval $I_i$ until interval $I_{i+1}$, or it may be *loaded* by delaying all packets from interval $I_{i-1}$ until interval $I_i$. A loaded interval will therefore have twice the expected number of packets, and a cleared one will have none. To send a 0 bit in position $i$, the interval $I_i$ is cleared and $I_{i+1}$ is loaded; to send a 1, $I_i$ is loaded and $I_{i+1}$ is cleared. (Note that since clearing one interval implicitly loads the next, it takes 3 intervals to send a bit.)

The watermarker and detector agree on the parameters $o$, $T$ and a list of positions $S = \{s_1, \ldots, s_n\}$; all of these parameters are secret. The watermarker encodes the watermark bits at the interval positions $s_i$ and the detector checks for the existence of the watermark. The check is performed by testing whether the data rate in interval $I_{s_i}$ differs from the rate in interval $I_{s_i+1}$ by a factor exceeding a threshold; if it does, then a 0 or 1 bit is considered detected. By focusing on data rates rather than individual packet timings, the interval-based approach is robust to repacketization of data flows.

The detection process may generate false positives due to natural variation in packet rates, or false negatives, as delays between the watermarker and repacketization at the relay cause rates in intervals to shift. To ensure reliable transmission, each watermark bit is encoded in several positions in the stream. Pyun et al. show that this technique operates with very low false-positive and false-negative rates.

### 2.6 Spread-Spectrum Watermarking

In DSSS watermarking technique of Yu et al. [8], each bit of a length-$n$ binary watermark is embedded in an interval of length $T_s$. Hence the whole watermark is inserted in some part of the flow of length $nT_s$. To embed a watermark bit 1, the rate of the packets in its length-$T_s$ designated interval are manipulated according to a Pseudo-Noise (PN) code. The PN code is a fast varying signal that switched between $+1$ and $-1$; the duration of each $\pm 1$ period is $T_c$. In particular, Yu et al. choose a length-7 PN code for their implementation. When PN code is $+1$, the rate of flow remains intact, but when PN code is $-1$, the rate of flow is decreased for a duration of $T_c$. The flow rate is manipulated by creating an interfering flow and relying on TCP congestion control. (Note that this approach works only with bulk flows where the sending rate is indeed limited by TCP congestion control.) On the other hand to embed a watermark bit 0, the flow is manipulated using the complement of the PN code.

The watermarker and detector agree on the parameter $T_s$, the watermark, and a Pseudo-Noise code. The detector recovers the watermark by first applying a high-pass filter to the received signal and subsequently passing it through despreading and a low-pass filter. The details of the detector's structure are inconsequential to our attack and the interested reader is referred to [8].

Given that the watermark insertion technique in DSSS reduces the flow rates over certain intervals across all flows it is vulnerable to our averaging attack, which is analysed in this paper. More recently, Huang et al. suggest to change the DSSS watermark to use different PN codes for watermarking different flows in order to defend against the multi-flow attack presented in this paper [21]. This approach results in increasing the false positive rates of the watermark detection as well as the complexity of the watermark detector, since a detector needs to correlate any received flow against all possible PN codes that might have been used for watermarking; unfortunately, this has not been considered by the authors.

## 3 Attack Analysis

In this section, we present a probabilistic analysis of our attack using a model for interactive traffic. Though some watermarked traffic may consist of non-interactive bulk transfer traffic, we will show in Section 4.1 that interactive traffic presents a more difficult case for our attack, and thus we analyze it here. As DSSS watermarks work well only against non-interactive traffic, our analysis here applies only to IBW and ICBW, but as we demonstrate experimentally, our attack will work on DSSS watermarks as well.

### 3.1 Probabilistic Model of Interactive Traffic

We first present a model for interactive traffic, as it is essential to our analysis. Let $f_m$ denote the $m$-th flow in a pool of interactive traffic flows. Given that the traffic might be encrypted, we do not consider the content of the packets; likewise, the sizes of packets representing keystrokes are likely to be uniform. We thus consider only the arrival time of the packets in the flow, allowing us to model the flow as a point process.

Suppose we observed packet arrivals at times $t_1 < t_2 < \cdots < t_n$ in a fixed interval $(0, \tau]$ such that $t_i$ is the time the $i$-th packet arrived. The collection of arrival times $\mathbf{t}_m = (t_1, t_2, \ldots, t_n)$ specifies a flow $f_i$. Furthermore, we model the interactive connection as a Markov-modulated Poisson process (MMPP) [22], [23]. The set of possible states are $\{0, 1\}$, where state 0 corresponds to user typing characters and state 1 corresponds to periods of silence. Figure 4 depicts this two-state MMPP.

Let $X(t)$ denote the state of the process at time $t$. When the process is at state 0, packet arrivals are modeled as a renewal process; i.e., the interarrival times are independent and identically distributed (i.i.d.). In case of interactive traffic flow this renewal process is often modeled as Poisson [19], [20]. The Poisson assumption means that the interarrival time of the packets, denoted by $\theta$, are exponentially distributed. Hence its probability density function (PDF) is given by:

$$f_\theta(t) = \lambda e^{-\lambda_0 t}$$

where $\lambda_0$ denotes the rate of the Poisson process. When the process is in state 1, the arrivals are again modeled as Poisson but with rate $\lambda_1 < \lambda_0$. Given that state 1 corresponds to a period of silence (no packet arrivals), as soon as a packet arrives the embedded Markov chain transitions to state 0. Therefore, the transition probabilities $\{P_{ij}, i \geq 0, j \geq 0\}$ of the embedded Markov chain $\{X_n, n \geq 0\}$ are as follows:

$$\begin{aligned} P_{00} + P_{01} = 1, \\ P_{01} = 1, P_{11} = 0 \end{aligned} \tag{1}$$

and the embedded Markov chain is defined by the matrix:

$$\begin{bmatrix} P_{00} & 1 \\ 1 - P_{00} & 0 \end{bmatrix}$$

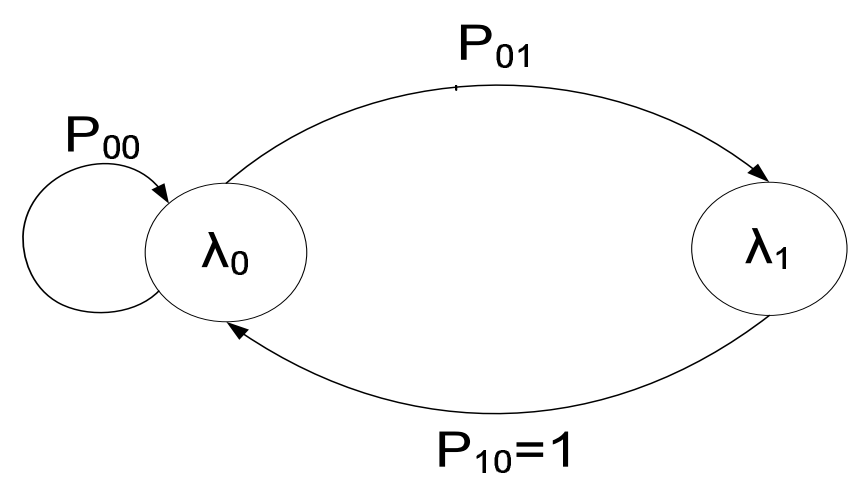

Fig. 4. The embedded two-state Markov chain.

The steady state probabilities $\pi_0, \pi_1$ of the embedded chain $X_n$ are given by:

$$\begin{bmatrix} \pi_0 \\ \pi_1 \end{bmatrix} = \begin{bmatrix} P_{00} & 1 \\ 1 - P_{00} & 0 \end{bmatrix} \begin{bmatrix} \pi_0 \\ \pi_1 \end{bmatrix}$$

or:

$$\pi_0 = \frac{1}{2 - P_{00}}, \qquad \pi_1 = \frac{1 - P_{00}}{2 - P_{00}}$$

The steady state probabilities $P_0, P_1$ of the Markov process $X(t)$ are given by ([23]):

$$P_i = \frac{\frac{\pi_i}{\lambda_i}}{\sum_k \frac{\pi_k}{\lambda_k}}$$

or:

$$P_0 = \frac{\lambda_1}{\lambda_1 + (1 - P_{00})\lambda_0}, \qquad P_1 = \frac{(1 - P_{00})\lambda_0}{\lambda_1 + (1 - P_{00})\lambda_0} \tag{2}$$

The significance of the steady state probabilities of (2) is that they capture the probability of each of the states 0 and 1 at any given point in time. Recall that ICBW encodes the watermark bits "1" or "0" by delaying the arrival times of the packets at the set of intervals $A_i$ or $B_i$ respectively and IBW encodes the watermark bits "0" or "1" by transferring the traffic of an interval of length $T$ to some adjacent interval. Therefore, they both creates periods of time with no arrivals in the flow. This period for ICBW is of length $a$ and for IBW is of length $T$. When the embedded Markov chain is in state $i$, we can compute the probability of zero occurring in a period of length $\ell$ starting at any given point as:

$$P_{f_m^i}(0; \ell) = e^{-\lambda_i \ell} \tag{3}$$

since the waiting times are exponentially distributed and therefore memoryless.

In general given a flow $f_m$ generated from an MMPP, from (3) probability of having a period of length $\ell$ with no arrivals $P_{f_m}(0; \ell)$ is:

$$\begin{aligned} P_{f_m}(0; \ell) &= P_0 P_{f_m^0}(0; \ell) + P_1 P_{f_m^1}(0; \ell) \\ &= P_0 e^{-\lambda_0 \ell} + P_1 e^{-\lambda_1 \ell} \end{aligned} \tag{4}$$

where the steady state probabilities $\{P_0, P_1\}$ are given by (2).

A good watermarking scheme requires that the watermarked stream should not reveal any clues of the presence of the watermark to unauthorized observer. Therefore, it is desirable that $P_{f_m}(0; \ell)$ above should be reasonably large so that presence of silent periods does not give away the watermark. We next present parameters of our two-state MMPP and show that for those parameters the watermark indeed cannot be detected with observing a single stream watermarked with ICBW or IBW. However, we will show that if the attackers have access to multiple copies of a marked signal, they can defeat the two watermarking schemes both when multiple flows are watermarked with the same key and when they are watermarked using various keys.

### 3.2 Parameter Selection and Goodness of Fit

We estimated the parameters $P_{00}$, $\lambda_0$, and $\lambda_1$ of our MMPP model by using network traces of SSH connections taken at a wireless access point in our institution. For a trace, we first estimated the underlying state of the embedded Markov chain by choice of a threshold $\eta$. If the interarrival time between two packets exceeded the threshold $\eta$, we assumed that the process was in state 1 and if the interarrival time between two packets was less than the threshold $\eta$, we assumed that the user was typing and therefore she/he was in state 0. Once the states $\{X_n, n \geq 0\}$ of the underlying chain are determined, by concatenation of the parts of the interactive traffic that came from same underlying state, we could extract two Poisson sub-flows with rates $\lambda_0$ and $\lambda_1$ from the original flow.

Given that the expected number of arrivals of a Poisson process distribution with parameter $\lambda$ in time interval $(0, t]$ is $\lambda t$, we estimated the rate $\lambda_0$ and $\lambda_1$ by calculating the arrival rates of each of the two extracted sub-flows. Parameter $P_{00}$ was estimated as the portion of the time the chain spent at state 0. Our estimated values for the transition probability $P_{00}$ and the rates $\lambda_0$ and $\lambda_1$ were as follows:

$$P_{00} = 0.96 \qquad \lambda_0 = 5.6 \qquad \lambda_1 = 0.57. \tag{5}$$

To assess the goodness of fit of our MMPP with parameters of (5), we used a quantile-quantile (q-q) plot [24]. Using the theoretical CDF of the model, the observations are mapped into values in interval $[0, 1]$. If the underlying statistical model of the data is consistent with the observations, the values obtained from the mapping are uniformly distributed in the interval $[0, 1]$. To assess the uniformity of the mapped values or equivalently assessing the goodness of the theoretical model an empirical CDF of the mapped values is compared against the theoretical CDF of a uniform distribution which is a 45-degree reference line. The closer the CDF to this reference line, the greater the evidence that the statistical model captures the underlying phenomenon. The q-q plot of Figure 5 for our model shows that the MMPP model for the interactive traffic with parameters (5) provides a good fit for the data and significantly outperforms a simpler Poisson model, or a Pareto distribution that has been previously proposed to fit interactive traffic [25].

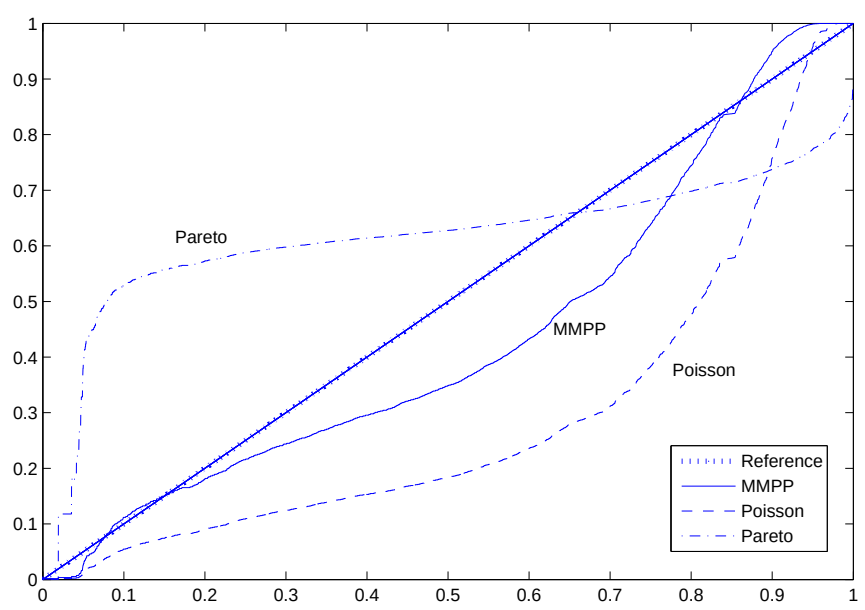

Fig. 5. Q-Q plot of Poisson and MMPP models with our sample data.

### 3.3 Multi-Flow Attack

Regardless of whether the ICBW or IBW watermarking schemes are implemented using the same message across all interactive flows or they use multiple message for different flows, they are subject to an averaging attack. This is because both schemes embed watermarks by emptying the same parts across various flows. Next, we will explain our attack for both the single-message and multiple-message watermarks.

#### *3.3.1 Averaging Attack against Single-Message Watermarks*

When ICBW or IBW watermarking schemes are implemented using the same message across all interactive flows, if the attacker has access to $k$ watermarked flows, he can form an aggregate of all the flows by taking the sorted union of all the arrival times of packets in all flows. We denote this aggregated stream by $\overline{f_k}$, where the subscript $k$ denotes the total number of streams involved in forming the aggregate flow.

Given that each interactive stream is independent of all the other streams, the probability of having a period of length $T$ with no arrivals in the flow $\overline{f_k}$ is given by:

$$\begin{aligned} P\{N_{\overline{f_k}}(t_a+\ell) - N_{\overline{f_k}}(t_a) = 0\} &= \prod_{i=1}^{k} P_{f_i}(0;\ell) \\ &= P_{f_m}(0;\ell)^k \quad (6) \end{aligned}$$

Equation (6) shows that probability of having period of length $\ell$ with no arrivals decreases exponentially in $k$, the total number of copies used to form the aggregate flow $\overline{f_k}$. Therefore, if the streams are not watermarked there is a very small probability that the aggregate stream has periods of no arrivals. However, if ICBW or IBW use the same key and message across all interactive flows, the aggregated copy of the watermarked flows always exhibits patterns of no arrivals of length $\ell$ that give away the location of the watermark as well as the maximum delay parameter $a$ of ICBW and the the period $T$ of IBW.

Substituting the parameters of (5) into (4), assuming $\ell = 350$ ms, as suggested by Wang et al. [7], we have $P_{f_m}(0; 0.35) = 0.33$. Therefore, in an aggregate of as few as 10 flows probability of a periods of 350 ms without any arrivals is as low as $P_{f_m}(0; 0.35)^{10} = 1.6 \times 10^{-5}$. Similarly for $\ell = 900$ ms, as used by Pyun et al. [6], we have $P_{f_m}(0; 0.9) = 0.17$ and $P_{f_m}(0; 0.9)^{10} = 2.4 \times 10^{-8}$.

This, of course, shows us the probability of finding an empty interval in a particular spot; we next consider the possibility of finding empty intervals at *any* position in the flows. To do so, we use a discrete approximation. Given an aggregate flow of length $L$, we are interested in finding the probability of having an empty interval of length $\ell$ at any position. For this, we divide the aggregate flow into non-overlapping intervals with length $\ell_M = \ell/M$ (a total of $N = \lfloor L/\ell_M \rfloor$ intervals). Finding $M$ (or $M-1$) consecutive empty intervals of length $\ell_M$ gives lowerbound (upperbound) of this probability.

Since $P_{00} = 0.96$, the process is nearly memoryless and we can approximate the discrete version of the problem as a Bernoulli process, where each interval is empty with probability $p_M = P_{\overline{f_k}}(0; L_M)$. For a total of $n$ intervals let us refer to the probability of finding $s$ consecutive empty intervals as $P_E(s, p_M, n)$. We can compute this using a recurrence.[2] Let $y[n] = P_E(s, p_M, n)^c$, i.e., the probability of finding *no* consecutive runs of $s$ empty intervals among the first $n$ intervals. Then, for $n \geq s$, we have:

2. This solution is adapted from [26].

$$y[n] = y[n-1] - (1-p_M)p_M^s \cdot y[n-s-1]$$

This is because the probability of having no runs of $s$ empty intervals among $n$ is the probability that there aren't any empty intervals among the first $n-1$, less the probability that there is exactly one run among the last $s$ intervals. This recurrence has the characteristic polynomial:

$$p(x) = x^{s+1} - x^s + (1-p_M)p_M^s$$

Any solution to the recurrence can be expressed in terms of the roots of the polynomial $p(x)$; given roots $r_i$ with respective multiplicities $m_i$, we have that:

$$y[n] = \sum_i p_i(n) r_i^n$$

where $p_i$ is a polynomial of degree at most $m_i - 1$. (See, for example, [27, Theorem 4.5.6].) Note that $y[n] = 1$ for $n < s$, which allows us to solve for the coefficients of the polynomials. Finally, we compute $P_E(s, p_M, n) = 1 - y[n]$.

Note that the schemes above will create *multiple* blank intervals, so we compute the probability of finding $e$ blank intervals of length $\ell$ in a flow of length $L$, $P'_E(L, \ell, e)$. Modeling the process as approximately memoryless, we can see that, for $e > 1$ and $L > \ell$:

$$P'_E(L, \ell, e) = P'_E(L, \ell, 1) P'_E(L-\ell, \ell, e-1)$$

Therefore:

$$P'_E(L, \ell, e) = \begin{cases} \prod_{i=0}^{e-1} P'_E(L - i\ell, \ell, 1) & L \geq e\ell \\ 0 & \text{otherwise} \end{cases}$$

where $P_E^{'}(L - i\ell, \ell, 1)$ is approximated by the method above, i.e.,

$$P_E(M-1, p_M, \lfloor (L-i\ell)/\ell_M \rfloor) \leq P_E^{'}(L - i\ell, \ell, 1) \leq P_E(M, p_M, \lfloor (L-i\ell)/\ell_M \rfloor)$$

We apply these computations to parameters, taken from the evaluation of the ICBW scheme by Wang et al. [7]. They used a 32-bit watermark, with a redundancy between 12 and 20, and flow lengths between 394 and 650 seconds. In Figure 6, we plot $P_{E'}(394\,\text{s}, 350\,\text{ms}, 12 \times 32)$ as a function of the number of aggregated flows. We can see that, even for small numbers of flows, the false-positive probability of our attack is quite low. This graph was computed using $M = 40$, which is sufficient to give an approximate error of less than $10^{-45}$ for $k > 4$, computed by comparing the upper and lower bounds.

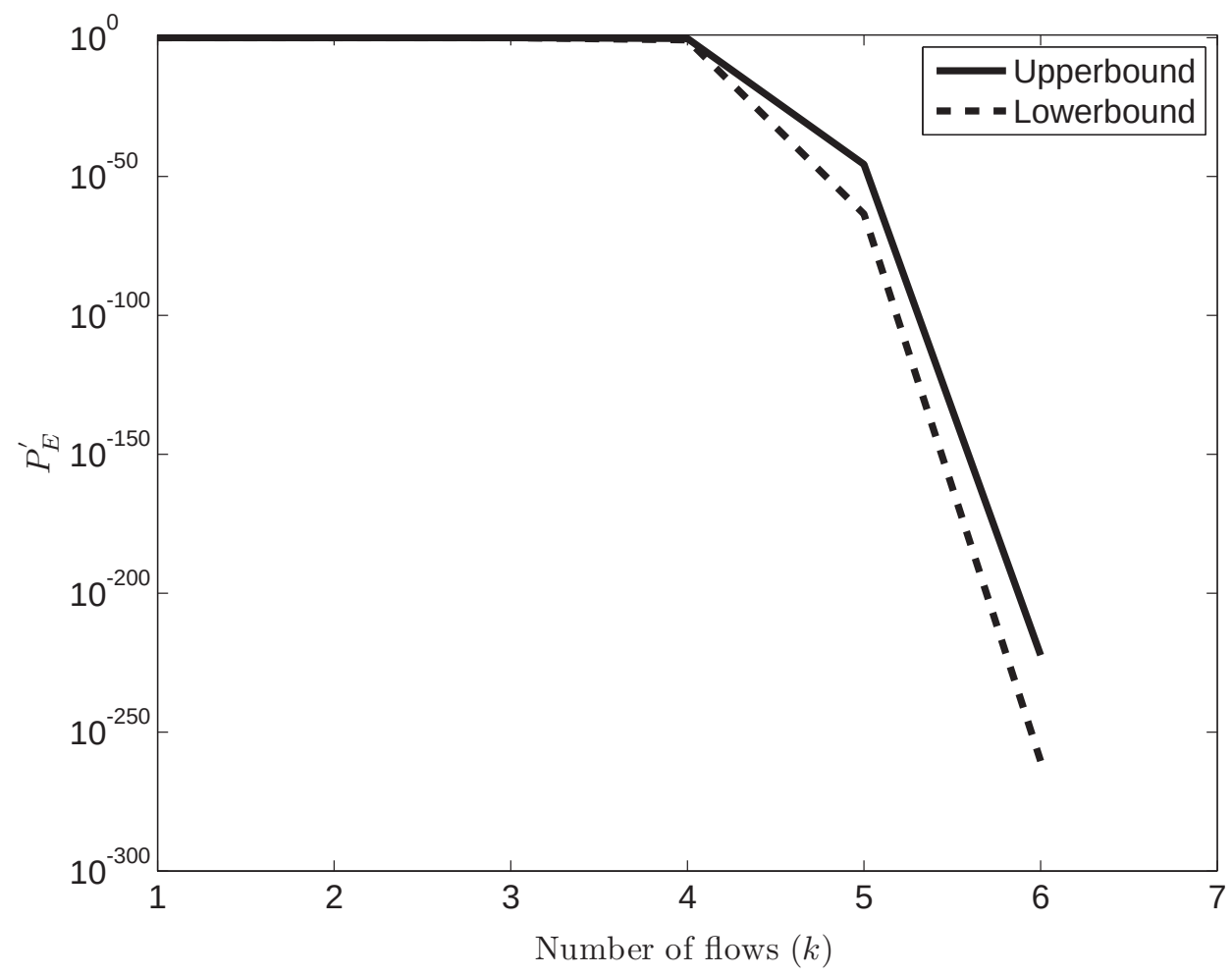

Fig. 6. False positive errors of MFA for different number of aggregated flows.

### 3.4 Impact of Timing Perturbations

Our analysis above assumed that the the attacker sees the timings of the watermarked stream directly. In reality, these timings will be perturbed by network delays. As a result, the intervals cleared by the watermark may have some packets from previous intervals shifted into them and no longer appear completely empty. Note that what is relevant here is not the magnitude of the network delay but its variance, or *jitter*, since delaying all packets by an equal amount does not affect our attack. And if the jitter is much less than $\ell$, our attack will work equally well: if jitter is $< \epsilon$ with high probability, then we will find clear intervals of length at least $\ell - \epsilon$ in the $k$ averaged watermarked streams, whereas the probability of seeing such an interval in unwatermarked streams is $P_{f_m}(0; \ell - \epsilon)^k \approx P_{f_m}(0; \ell)^k$, which is vanishingly small. We observe that the studied parameters of the ICBW and IBW schemes have $\ell = 350\,\text{ms}$ or $900\,\text{ms}$, in order to resist traffic perturbations, repacketization, etc. The network jitter, on the other hand, is two orders of magnitude smaller. Our experiments on PlanetLab [28] show it to be on the order of several milliseconds for geographically distributed hosts, and this matches the results of previous studies [29]. Therefore, it is indeed the case that the jitter is $< \epsilon \ll \ell$, and so it will not significantly affect our attack.

## 4 IMPLEMENTATION

Having shown the theoretical background behind our attack, we now show the result of implementing it in practice. We developed algorithms to detect the presence of a watermark, recover the secret parameters, and to remove the watermark from new streams. We evaluated the algorithms using both real flows gathered from traces and synthetic flows generated using our MMPP model, presented in Section 3.1. We first present our attacks for same-value watermarks, and then extend it to multi-valued watermarks.

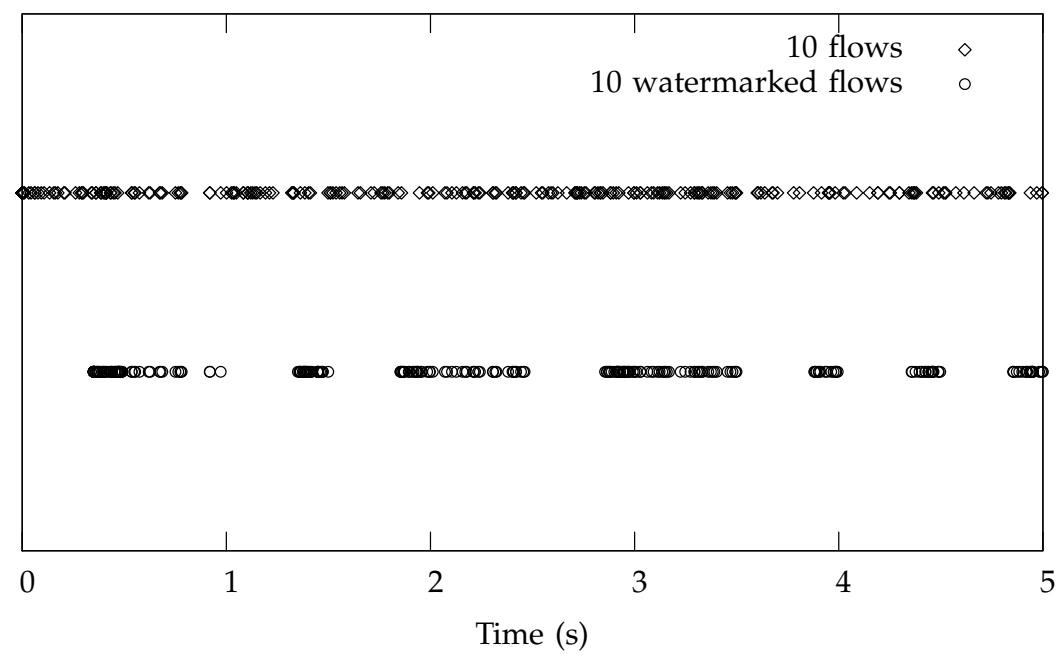

(a) Interval-Centroid Based Watermark

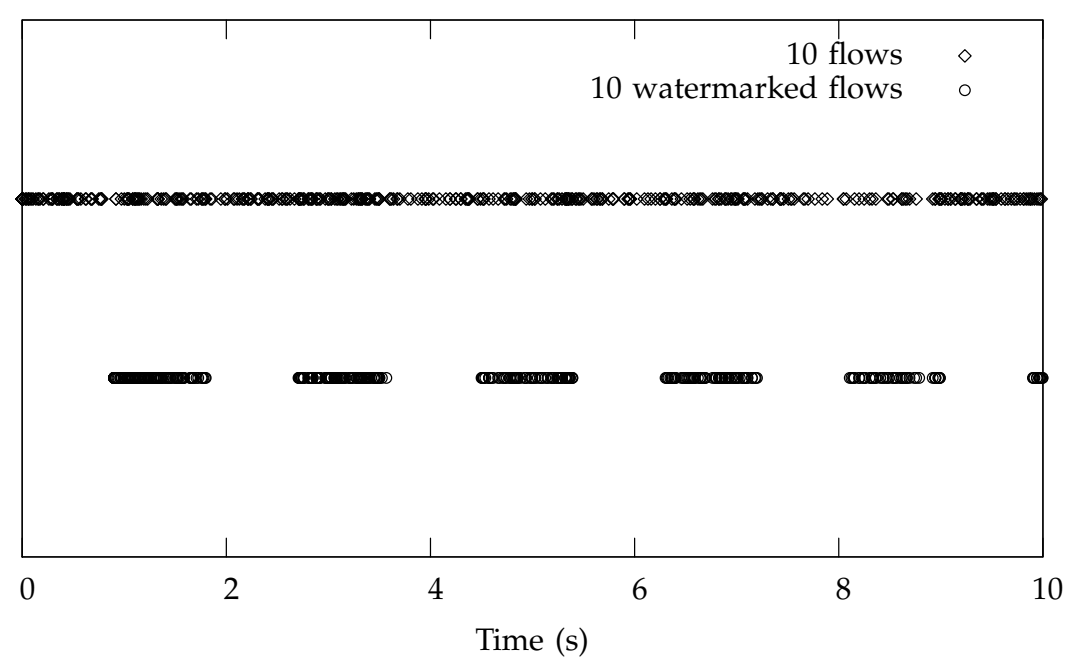

(b) Interval-Based Watermark

Fig. 7. 10 flows before and after watermarking.

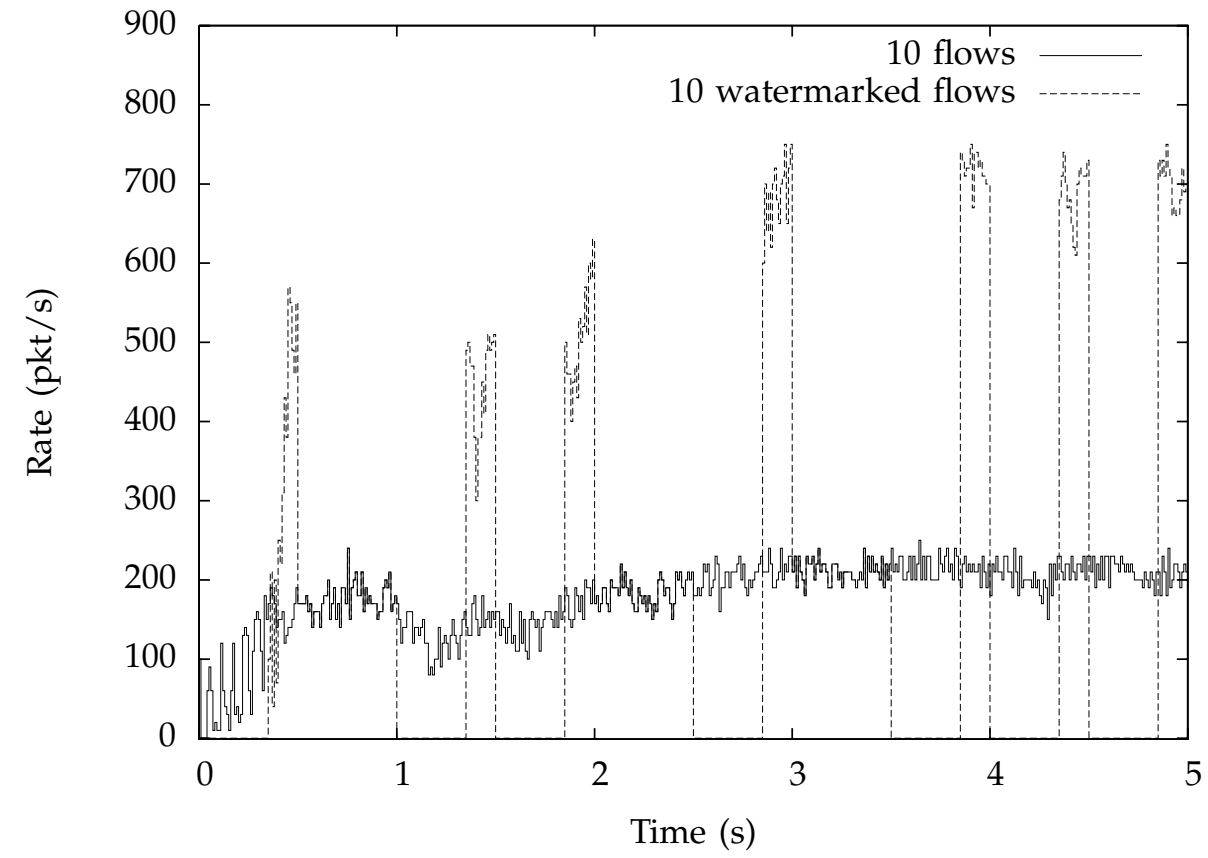

(a) Watermark on 10 flows

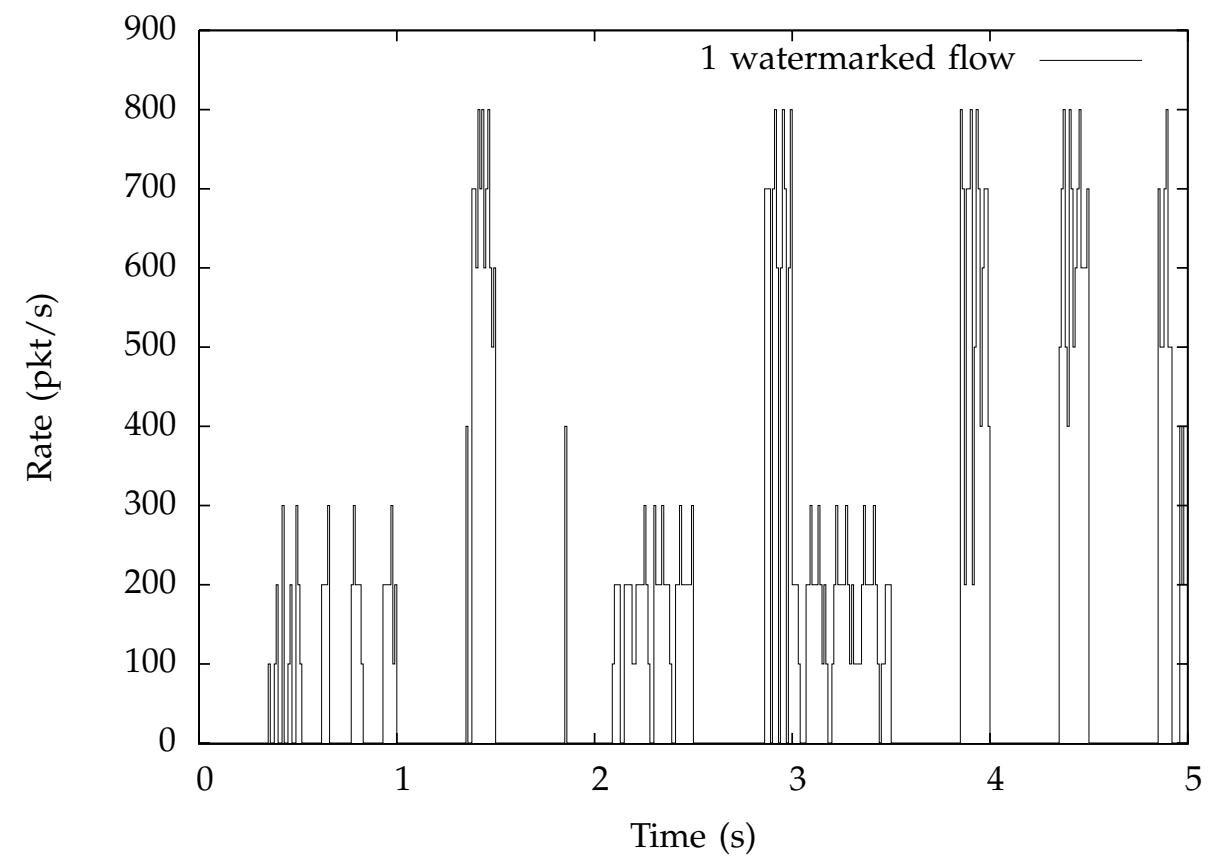

(b) Watermark on 1 flow

Fig. 8. Watermark detection on bulk traffic.

### 4.1 Watermark Detection

As above, our attack relies on collecting a series of flows that are watermarked with the same value. These flows are combined into a single flow and examined for large gaps between packets. Figure 7(a) shows the packet arrivals for 10 combined flows before and after an ICBW watermark has been applied. The watermark pattern is clearly visible in the combined flows, alerting about the watermark presence. Figure 7(b) shows the same process working with the IBW watermark scheme.

We also performed the same analysis for non-interactive, bulk transfer traffic by applying the watermark to packet traces we collected from web downloads across a DSL connection. Figure 8(a) shows the packet timings for 10 combined flows before and after a watermark. Bulk transfers have a somewhat more regular behavior, since they are controlled by the TCP algorithms, rather than by individual users. This can be seen at the beginning of the 10 combined flows before watermark: the TCP slow start period results in a much lower rates for the first few seconds of the connection. However, this regularity quickly gets out of sync due to irregular network delay and response times. In the graph of 10 watermarked flows, the intervals squeezed by the watermark are readily visible. In fact, because data transfer flows are much more dense than interactive flows, the watermark is visible even on a single flow (Figure 8(b)).

The DSSS watermark is intended to be applied to bulk transfer traffic such as FTP, since it interferes with traffic rate, rather than changing packet timings. A similar muiti-flow attack works against DSSS as well, as shown in Figure 9. (We used the parameters of chip length 0.4s, chip sequence length of 7, and code length of 7.) In this case, periods of high interference are clearly seen as low-rate periods in the flows, allowing one to recover the chip sequence and then decode the watermark.

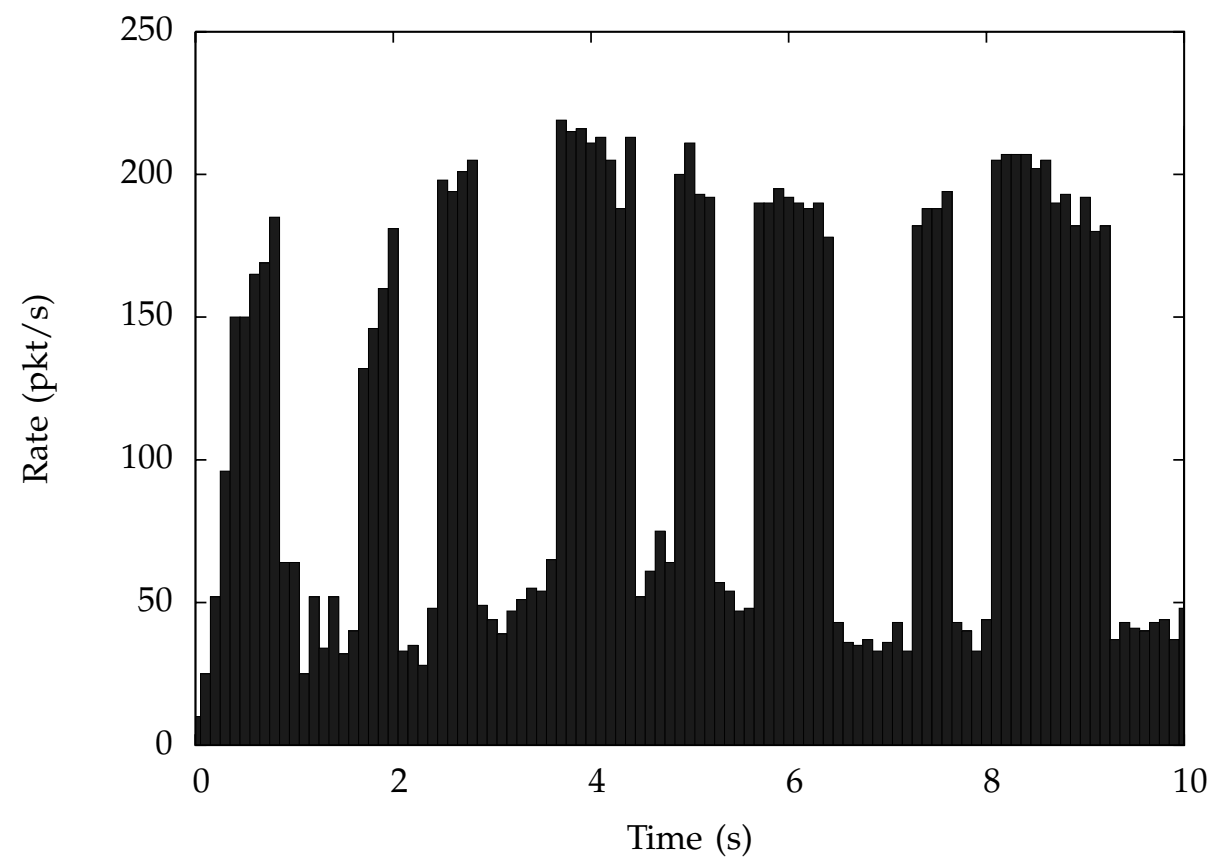

Fig. 9. Average rate of 10 flows after DSSS watermark.

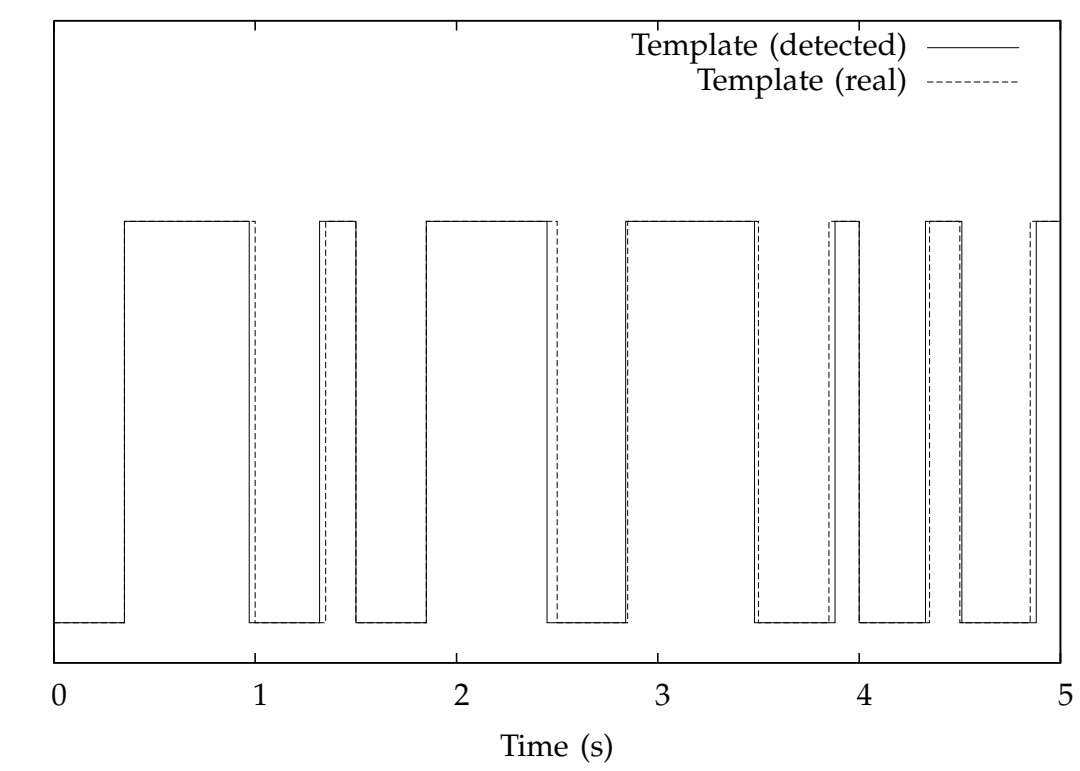

(a) Template of clear intervals

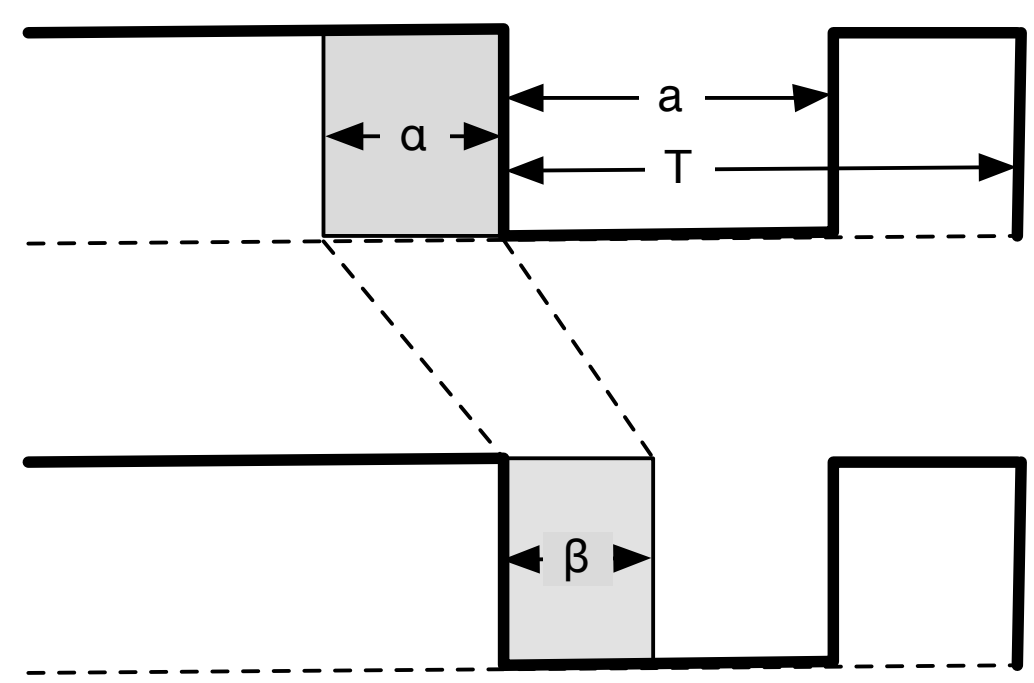

(b) Shift to remove watermark

Fig. 10. Watermark Removal

### 4.2 Watermark Removal

Based on the combined graphs, it is easy to recover the watermark parameters as well. We can build a template of clear intervals by selecting all intervals larger than a threshold; for example, Figure 10(a) shows the template derived from 10 flows watermarked by ICBW. The estimated template is somewhat imprecise, due to network jitter, as well as the fact that small (10–20ms) gaps may precede or follow the clear intervals even when 10 flows are combined. However, this imprecision is not a problem since the watermark can still be effectively removed. The template also lets us estimate the values of $T$ and $a$. We can average the lengths of clear intervals and the distance between two consecutive clear intervals to obtain a relatively precise estimate. Armed with this information, we can then modify a new flow to remove the watermark.

For ICBW, we have two choices: we can either shift traffic into the clear intervals in the template, thereby negating the squeezing action of the watermark, or find intervals that have not been squeezed and squeeze them. We decided to implement the former approach since it does not require as precise an estimate of $T$. Also, it leaves the flow looking more natural. Our shift is implemented as shown in Figure 10(b), by shifting all packets in a period $\alpha$ before the clear interval into an interval of length $\beta$ inside the clear interval. Larger values of $\alpha$ and smaller values of $\beta$ will more significantly shift the interval centroid back in a different direction; however, very small values of $\beta$ may not have the desired effect, since the template is imprecise and too many packets may get shifted without arriving into the correct interval. Experimentally, we found that $\alpha = 0.9(\hat{T} - \hat{a})$ and $\beta = 0.8(\hat{T} - \hat{a})$ provides best results, where $\hat{T}$ and $\hat{a}$ are estimated values of $T$ and $a$.

Table 1 shows the results of watermark removal. We reimplemented the ICBW detection mechanism and computed the Hamming distance of the encoded watermark to the detected one, collected over 100 flows. (We show the average distance, with range shown in parentheses). With as few as 10 flows, we are able to get a reasonably good estimate of $T$ and $a$ and remove the watermark in most cases—the ICBW detection scheme uses a Hamming distance threshold of 5–8 to decide when a watermark has been detected. With 15 flows, we get a more accurate template and estimate, and all 100 flows will clear the template.

A similar approach can be used to attack the IBW watermark; by delaying packets so that they fall into the clear intervals, the clear intervals become indistinguishable from loaded ones. Table 2 shows the effect of applying our attack on the IBW watermark, where 24 bits are encoded at different levels of redundancy. Even with a redundancy of 80, most bits are not recovered correctly. These results were obtained by using the code provided by the authors of [6].

TABLE 1
Results for removing ICBW watermarks

| Num flows | $\hat{T}$ | $\hat{a}$ | Hamming not watermarked | Hamming watermarked | Hamming attacked | Ave. delay | Max delay |
|---|---|---|---|---|---|---|---|
| 10 | 365 ($\sigma = 10.7$) | 492 ($\sigma = 15.2$) | 17.9 (13–24) | 2.67 (1–7) | 13.9 (2–20) | 33.6 | 164 |
| 15 | 353 ($\sigma = 0.60$) | 504 ($\sigma = 1.62$) | 17.6 (13–25) | 2.74 (0–6) | 16.1 (12–21) | 42.6 | 188.2 |
| 20 | 346 ($\sigma = 0.30$) | 504 ($\sigma = 0.50$) | 17.2 (12–21) | 2.68 (0–5) | 16.4 (11–20) | 45.4 | 194.3 |

We expect a similar technique should work against DSSS watermarks; a template of low rates can be inferred from several flows. An attacker can then decrease rates in the non-interference section of the template by dropping packets, or increase the rate in the high-interference section by delaying packets into the template. We do not have experimental results for DSSS since the detection algorithm is fairly complex and we did not have access to an implementation of it.

TABLE 2
Watermark bits detected before and after applying the attack (watermark length is 24).

| Rep. | Bits detected | | Marked packets |
|---|---|---|---|
| | Before attack | After attack | |
| 1 | 7 | 3 | 53 |
| 5 | 14 | 5 | 156 |
| 10 | 24 | 4 | 505 |
| 15 | 24 | 2 | 754 |
| 20 | 24 | 2 | 967 |
| 24 | 24 | 2 | 1209 |
| 30 | 24 | 2 | 1440 |
| 35 | 24 | 2 | 1724 |
| 41 | 24 | 2 | 2008 |
| 45 | 24 | 2 | 2307 |
| 50 | 24 | 2 | 2697 |
| 55 | 24 | 2 | 3083 |
| 60 | 24 | 2 | 3296 |
| 65 | 24 | 2 | 3623 |
| 70 | 24 | 2 | 3876 |
| 75 | 24 | 2 | 4090 |
| 80 | 24 | 2 | 4343 |

### 4.3 Multiple Values

So far we have assumed that the watermarks on all of the aggregated flows are the same. Here, we consider the case where each watermark uses multiple, different values. We can still execute our attack by relying on the fact that within a collection of $2k-1$ flows, for any given bit $b$, we can find $k$ flows where this bit has the same value (we have further discussed this in [11] and [30]).

Figure 11(a) plots the result of such a subset search. By inspection, we can see that in the first subset of flows, the interval (4.5,4.85) has been cleared. In the second subset, this interval remains cleared and the interval (0,0.35) becomes clear as well. The third subset has no packets in (2.0,2.35) and the fourth in (3.5,3.85). Note that this pattern immediately lets us detect the presence of a watermark; Figure 11(b) shows the same flow subsets on an unwatermarked section.

Recovery of the secret parameters can proceed largely as in the single-value case. One difficulty is that with the flow subsets, we may encounter large intervals that are not precisely aligned with the interval positions. For example, Table 3(a) lists the blank intervals longer than 0.2s in the last subset. There are a lot of wrong-size intervals that result from the case when 8 or 9 of the flows in the subset have had an interval squeezed, but the last one or two add a few packets to the mix. To address this concern, we can select the largest empty intervals in any subset, as shown in Table 3(b). These will correspond to intervals that have been squeezed on every flow. This can be used to recover the watermark parameters of $T$ and $a$.

Once these are obtained, the next step is to scan through all subsets and determine which intervals are always squeezed at the same time and call such lists $S_i$; these will correspond to either $A_b$ or $B_b$ for

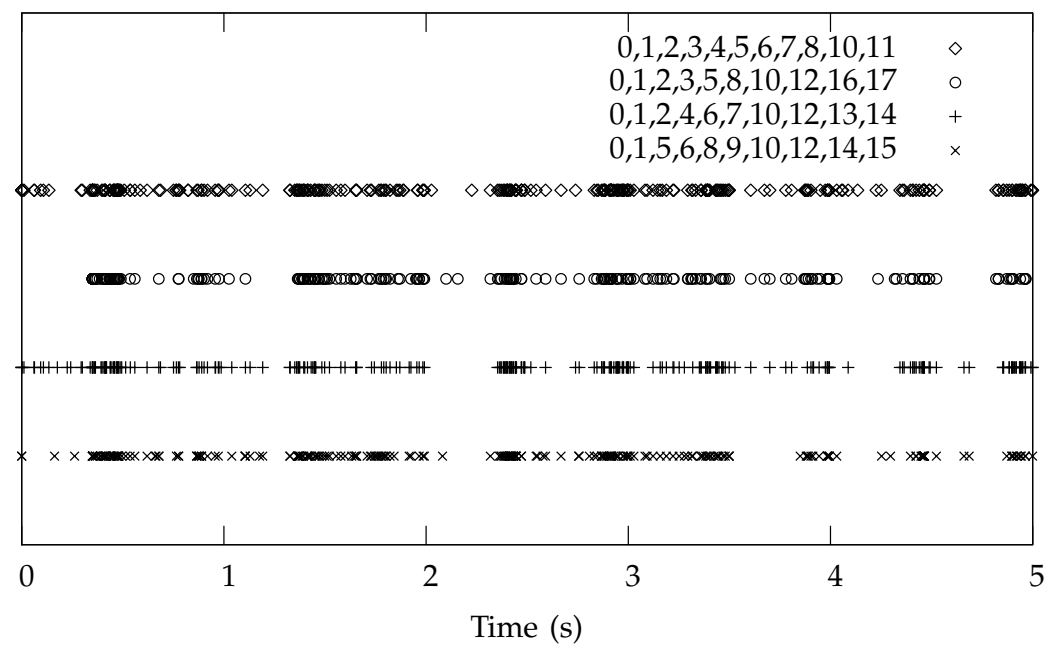

(a) Watermarked flow subsets

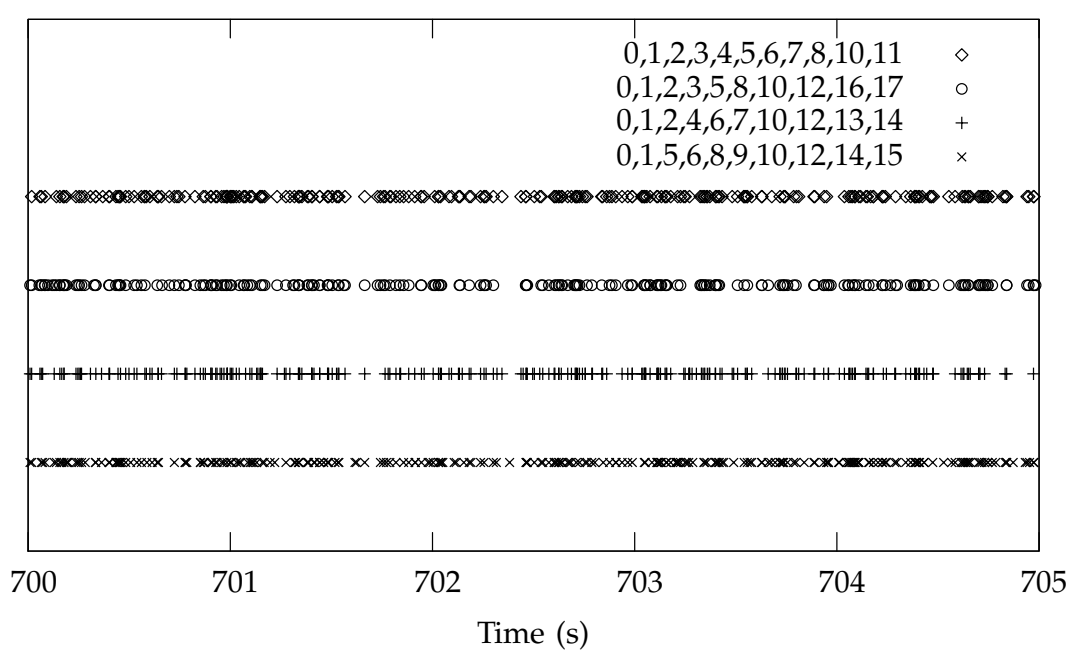

(b) Un-watermarked flow subsets

some bit $b$. Then, for each $S_i$, we find $S_j$ such that $S_i$ and $S_j$ are never squeezed at the same time. This will tell us that $S_i$ and $S_j$ correspond to the same bit. Armed with this knowledge, we can remove the watermark by observing the watermarked stream for a short while, and when we see intervals from $S_i$ that are being squeezed, we proceed to artificially squeeze intervals in $S_j$ (or unsqueeze further intervals in $S_i$, or both).

## 5 COUNTERMEASURES

We next consider several countermeasures to our attack.

### 5.1 Multiple Offsets

A watermark can be inserted at an offset $o$ from the start of the stream. This offset is picked randomly from the range $[0, o_{max}]$; [6] suggested to use $o_{max} = T$. An offset watermark can still be detected by enumerating different offsets and choosing the one with the highest detection result. This will increase the false positives, in proportion to $o_{max}$, but overall [6] reports that such a scheme still has good performance.

Since an offset is chosen randomly for each stream, it complicates the multi-flow attack because the watermark insertion points no longer line up

TABLE 3
Blank intervals from subset of flows

(a) All blank intervals

| Start | End |
|---|---|
| 2.08 | 2.32 |
| 3.50 | 3.85 |
| 4.03 | 4.25 |
| 5.13 | 5.33 |
| 11.59 | 11.85 |
| 18.14 | 18.37 |
| 19.56 | 19.79 |
| 25.58 | 25.82 |
| 30.06 | 30.34 |
| 34.08 | 34.35 |
| ... | ... |

(b) Largest blank intervals

| Start | End |
|---|---|
| 130.98 | 131.35 |
| 140.49 | 140.86 |
| 151.99 | 152.36 |
| 161.99 | 162.35 |
| 235.99 | 236.37 |
| 306.49 | 306.86 |
| 334.49 | 334.86 |
| 368.49 | 368.86 |
| 43.99 | 44.36 |
| 51.98 | 52.35 |
| ... | ... |

with one another. It becomes necessary to search for optimal alignments by trying multiple offsets for different streams. A simple approach is to select a step value $\delta$ and choose offset values from: $(0, \delta, 2\delta, ..., \lceil o_{max}/\delta \rceil \delta)$. The attacker will need to enumerate through each of these values for each stream out of $k$, evaluating $(\lceil o_{max}/\delta \rceil + 1)^k$ possibilities in all.[3]

Each target alignment might be imperfect, but it is easy to see that, for some choice of offset for each stream, the misalignment will be bounded by $\delta/2$. Therefore, we must search for clear intervals of length $\ell = T - \delta/2$. We can therefore bound the probability of false positives in the overall process by:

$$P_{FP} \leq \left( \left\lceil \frac{o_{max}}{\delta} \right\rceil + 1 \right)^k P_E^{'}(L, \ell, e) \qquad (7)$$

where $L$ is the maximum length of the streams and $e$ is number of required empty intervals for watermark detection ($P_E^{'}(L, \ell, e)$ is analyzed in Section 3.3).

Figure 11 illustrates the corresponding false positive error rate for different number of flows $k$ when the maximum offset value is $o_{max} = 10T$ and the step value is $\delta = T$. Comparing with using only a single offset (Figure 6), we can see that the multi-flow attack is still effective, at the cost of more computation for the attacker and requiring more

3. The computational requirements can be reduced by eliminating from consideration any combinations that can be shown to lack the necessary clear intervals in a subset of all streams. E.g, if the first two streams have no intersecting clear intervals that are long enough with offsets $(0, 0)$, it is not necessary to consider combinations with other stream at offsets $(0, 0, \ldots)$.

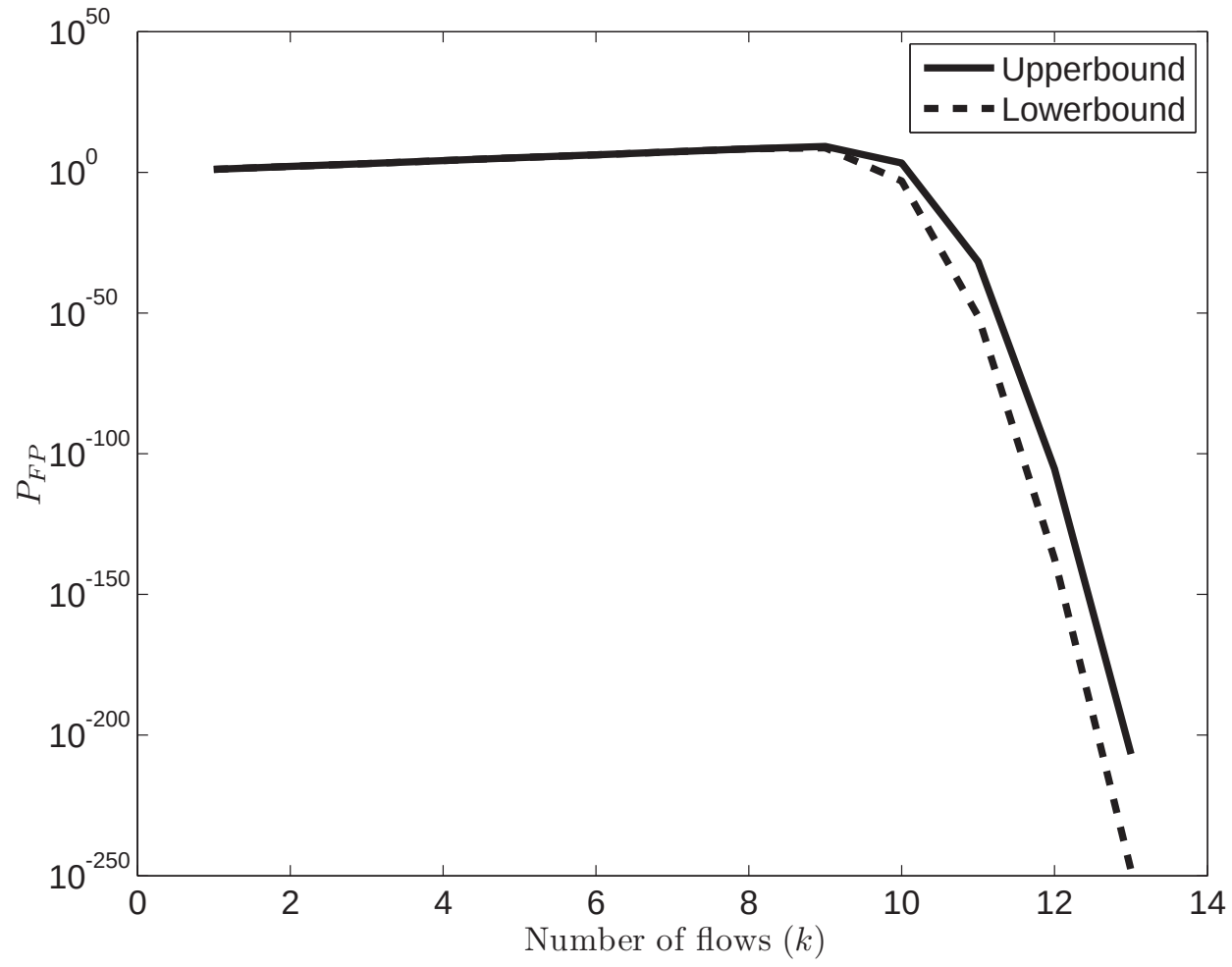

Fig. 11. False positive errors of MFA for different number of aggregated flows when multiple offsets are used ($o_{max} = 10T$, $\delta = T$).

(approximately twice) watermarked flows for the same performance. It should also be mentioned that this also increases the false positive of the watermark detector by a factor of $o_{max}/\delta = 10$. Note that larger $o_{max}$ increases the attacker's false positive and also the computation, but requires longer flows to insert the watermark.

### 5.2 Multiple Positions

Another alternative is to choose different positions, in the case of ICBW and IBW, and different PN codes in the case of DSSS [30]. Let us consider the case of ICBW. A watermarker and detector must use the same assignment of intervals to the sets $A_i$ and $B_i$, as determined by the random seed $s$, in order for the watermark to be successfully recovered. However, a watermarker may decide to use multiple seed values, $s_1, \ldots, s_n$, and pick one of them at random for each flow.

To deal with this, the detector would need to try to recover the watermark with each possible $s_i$ and pick the best match. Once again, the probability of error grows with $n$, but increased redundancy can again be used to make up for it. Note that the probability of error falls exponentially with increased redundancy, but grows only roughly linearly with $n$.

We can once again use the subset attack to try to find $k$ flows that use the same seed value $s_i$; however, the complexity grows quickly out of control. The probability of a given set of $k$ flows using the same seed is $\left(\frac{1}{n}\right)^{k-1}$, which falls quite quickly even when $k = 10$. By the pigeon hole principle, within $n(k-1)+1$ flows we can always find a subset of $k$ flows with the same seed, but the search space of all $\binom{n(k-1)+1}{k}$ subsets grows superexponentially in $n$. For example, with $n = 6$ and $k = 10$, $\binom{51}{10} > 10^{10}$, resulting in an infeasible number of subsets to enumerate.

The same principle can apply to IBW, by picking multiple sets of positions $\{s_i\}$, and to DSSS by using multiple PN codes [21].

## 6 Conclusion

We have demonstrated an attack on three recent network flow watermarking schemes that is highly successful, while requiring a low amount of resources. Our attack, MFA, is based on a solid theoretical grounding, and has been validated with a prototype implementation tested against the original prototypes. MFA can detect the presence of the watermark on a watermarked flow and remove it successfully. Additionally, in case of IBW scheme we can also recover the watermark parameters and values, allowing us to modify the watermark or insert it into other streams, confusing the detector. We have also suggested two countermeasures to our attack — switching bit positions and using different offset values. These countermeasures can impose a very high computation cost and therefore disable the attack.

While the use of network flow watermarking techniques for various security applications is quite new [4], [6], [7], [8], [9], [10], digital watermarking and specifically multimedia watermarking is a nearly mature field. Indeed most of network flow watermarking schemes are inspired by multimedia watermarks. To name a few Wang and Reeves's [4] scheme is a special instance of QIM watermarking, a well-understood multimedia watermarking technique [31]. IBW scheme of Pyun et al. [6] that we have broken is based on patchwork watermark of Bender et al. [32] and the scheme of Yu et al. [8] is based on spread spectrum watermarking [33].

The current approach for designing network flow watermarks suffers from the fact that while watermarking schemes are inspired by the digital watermarking schemes, little attention is given to the entirety of the watermarking design problem. For example, statistical characteristics of the underlying media are always an important consideration in digital watermarks, but network watermark research does not adequately model the effect that network traffic characteristics have on watermarks;

as we showed, the density of bulk traffic makes it very difficult to insert a transparent watermark. Likewise, digital watermarks have long considered the possibility that multiple watermarked documents can be used to attack watermarks [33], [34], but we are unaware of previous work looking at the multi-flow threat model for watermarking. We thus hope that future work on watermarks will be informed by our work and perform a broader analysis.